\documentclass[aps]{revtex4}%
\usepackage{amssymb}
\usepackage{amsfonts}
\usepackage{amsmath}
\usepackage{graphicx}%
\providecommand{\U}[1]{\protect\rule{.1in}{.1in}}

\begin{document}
\title{Low-Lying Collective Excitations of Superconductors and Charged Superfluids}
\author{S. N. Klimin}
\affiliation{{TQC, Universiteit Antwerpen, Universiteitsplein 1, B-2610 Antwerpen, Belgium}}
\thanks{Author to whom any correspondence should be addressed. E-mail:
\href{mailto:sergei.klimin@uantwerpen.be}{sergei.klimin@uantwerpen.be}}
\author{J. Tempere}
\affiliation{{TQC, Universiteit Antwerpen, Universiteitsplein 1, B-2610 Antwerpen, Belgium}}
\altaffiliation{Also at: {Lyman Laboratory of Physics, Harvard University, USA}}

\author{H. Kurkjian}
\affiliation{Laboratoire de Physique Th\'{e}orique, Universit\'{e} de Toulouse, CNRS, UPS,
31400, Toulouse, France}

\begin{abstract}
We investigate theoretically the momentum-dependent frequency and damping of
low-lying collective excitations of superconductors and charged superfluids in
the BCS--BEC crossover regime. The study is based on the Gaussian
pair-and-density fluctuation method for the propagator of Gaussian
fluctuations of the pair and density fields. Eigenfrequencies and damping
rates are determined in a mutually consistent nonperturbative way as complex
poles of the fluctuation propagator. Particular attention is paid to new
features with respect to preceding theoretical studies, which were devoted to
collective excitations of superconductors in the far BCS regime. We find that
at a sufficiently strong coupling, new branches of collective excitations
appear, which manifest different behavior as functions of the momentum and the temperature.

\end{abstract}
\date{June 23, 2023}
\maketitle

\section{Introduction}

Superconductors and fermionic superfluids exhibit a rich variety of collective
excitations, which are provided by the density and pair field response. The
majority of known superconductors are realized at high concentrations of
electrons such that the plasma energy is large with respect to the gap,
$\hbar\omega_{p}\gg\Delta$. Under this condition, the response of
superconductors and charged superfluids in the frequency range $\omega
\sim\omega_{p}$ is dominated by the plasma branch of collective excitations
\cite{Takada1998}. At lower frequencies, namely in the range of the
pair-breaking threshold, other collective excitations can be observable in
superconductors and charged superfluid systems, e.g., pair-breaking Higgs
modes~\cite{Shimano} and gapless Carlson--Goldman modes~\cite{Carlson}. The
low-lying collective excitations can be resolved experimentally in both
neutral and charged Fermi superfluids using different methods, such as Bragg
spectroscopy~\cite{Hoinka}, spatially resolved interferometry~\cite{Carusotto}%
,~and pump-THz probe spectroscopy~\cite{MatsunagaPRL}.

The low-lying collective excitations are well studied theoretically for
superconductors~\cite{Takada1997,Shimano} in the weak-coupling BCS regime.
Recent progress in a realization of the BCS--BEC crossover regime in
superconductors makes the treatment of collective excitations in the whole
crossover range timely and relevant.

We consider collective excitations as small Gaussian fluctuations on the top
of a uniform mean-field solution for the pair and density fields. This
formalism is equivalent to the random phase approximation (RPA)
\cite{Anderson1958}. Using these methods, collective excitations in charged
superfluids have been investigated~\cite{Plasmons,Plasmons2}, being
concentrated on the case of relatively small plasma frequencies, which can be
in resonance with the pair-breaking threshold. The present study is devoted to
the other regime, when the plasma frequency is large with respect to the gap,
which is relevant for existing superconductors. For $\hbar\omega_{p}\gg\Delta
$, the plasma branch of the response is only slightly affected by the
difference between the BCS--BEC crossover and the far BCS regime. Therefore,
the paper is particularly focused on the dispersion and damping of low-lying
collective excitations.

\section{Gaussian Pair-and-Density Fluctuation Method}

The present treatment follows the theoretical scheme which is represented in
detail in Ref.~\cite{Plasmons2}. Therefore, we only briefly describe the main
steps of the derivation. It exploits the partition function of a gas of
fermions with the spin projection $\sigma=\pm1/2$, which is a path integral
over anticommuting Grassmann fields $\left\{  \bar{\psi}_{\sigma},\psi
_{\sigma}\right\}  $,%
\begin{equation}
\mathcal{Z}=\int e^{-S}D\left[  \bar{\psi},\psi\right]  \label{PF}%
\end{equation}
with the action {functional}
\begin{align}
S  &  =\int_{0}^{\beta}d\tau\int d\mathbf{r}\left[  \sum_{\sigma
=\uparrow,\downarrow}\bar{\psi}_{\sigma}\left(  \frac{\partial}{\partial\tau
}+H-\mu\right)  \psi_{\sigma}+g\bar{\psi}_{\uparrow}\bar{\psi}_{\downarrow
}\psi_{\downarrow}\psi_{\uparrow}\right] \nonumber\\
&  +\frac{1}{2}\int_{0}^{\beta}d\tau\int d\mathbf{r}\int d\mathbf{r}^{\prime
}\frac{e^{2}}{4\pi\epsilon_{0}\varepsilon\left\vert \mathbf{r}^{\prime
}-\mathbf{r}\right\vert }\sum_{\sigma,\sigma^{\prime}}\bar{\psi}_{\sigma
}\left(  \mathbf{r}\right)  \bar{\psi}_{\sigma^{\prime}}\left(  \mathbf{r}%
^{\prime}\right)  \psi_{\sigma^{\prime}}\left(  \mathbf{r}^{\prime}\right)
\psi_{\sigma}\left(  \mathbf{r}\right)  . \label{S}%
\end{align}
The interparticle interaction in (\ref{S}) is a sum of the attractive contact
potential with the coupling constant $g<0$ and the repulsive Coulomb
potential, in which $\epsilon_{0}$ is the permittivity of free space and
$\varepsilon$ is the high-frequency dielectric constant of a medium.
Additionally, $\beta=\hbar/k_{B}T$ is the inverse temperature, and $\mu$ is
the chemical potential. We use the set of units with $\hbar=1$, the particle
mass $m=1/2$, and the particle density $n$ such that the Fermi wave vector of
free fermions in 3D is $k_{F}\equiv\left(  3\pi^{2}n\right)  ^{1/3}=1$.
Consequently, the free-particle Fermi energy $E_{F}$ in these units is
$E_{F}=1$.

Next, we obtain the effective bosonic action through standard steps as
described in detail in Ref.~\cite{Plasmons2}. First, we introduce the
auxiliary bosonic fields: the pair field $\left[  \bar{\Psi},\Psi\right]  $
and the density field $\left[  \Phi\right]  $. Second, the
Hubbard--Stratonovich shift of bosonic variables allows us to perform the path
integration over fermionic variables analytically exactly. This results in the
effective bosonic action%
\begin{equation}
S_{eff}=-\operatorname*{tr}\left[  \ln\left(  -\mathbb{G}^{-1}\right)
\right]  +\int_{0}^{\beta}d\tau\int d\mathbf{r}\left(  -\frac{1}{g}\bar{\Psi
}\Psi+\frac{1}{8\pi}\left(  \nabla\Phi\right)  ^{2}\right)  . \label{Seff1}%
\end{equation}
with the inverse Nambu matrix%
\begin{equation}
-\mathbb{G}^{-1}=\left(
\begin{array}
[c]{cc}%
\frac{\partial}{\partial\tau}+H-\mu+i\sqrt{\alpha_{0}}\Phi & -\Psi\\
-\bar{\Psi} & \frac{\partial}{\partial\tau}-H+\mu-i\sqrt{\alpha_{0}}\Phi
\end{array}
\right)  . \label{NambuMatrix}%
\end{equation}
{Here,} $\alpha_{0}$ is the dimensionless coupling strength for the Coulomb
interaction,%
\begin{equation}
\alpha_{0}\equiv\frac{e^{2}}{4\pi\epsilon_{0}\varepsilon\hbar}\sqrt{\frac
{2m}{E_{F}}}.
\end{equation}
{The} relation between $\alpha_{0}$ and the bare plasma frequency $\omega
_{p}=\sqrt{e^{2}n/\epsilon_{0}\varepsilon m}$ in the chosen units is given by
$\omega_{p}=\sqrt{\left(  8/3\pi\right)  \alpha_{0}}$.

The expansion of the effective bosonic action (\ref{Seff1}) up to the
quadratic order in powers of fluctuations of the pair and density fields about
the mean-field solution leads to the Gaussian pair-and-density fluctuation
(GPDF) action. We approximate the mean-field solution by uniform values,
neglecting exchange scattering contributions~\cite{Plasmons}. Consequently,
the gap equation remains the same as for a neutral superfluid:%
\begin{equation}
\int_{k<K_{0}}\frac{d\mathbf{k}}{\left(  2\pi\right)  ^{3}}\left(  \frac
{1}{2E_{\mathbf{k}}}\tanh\left(  \frac{\beta E_{\mathbf{k}}}{2}\right)
-\frac{1}{2k^{2}}\right)  +\frac{1}{8\pi a_{s}}=0, \label{gap2}%
\end{equation}
where $E_{\mathbf{k}}=\sqrt{\xi_{\mathbf{k}}^{2}+\Delta^{2}}$ is the pair
excitation energy, $\xi_{\mathbf{k}}=k^{2}-\mu$ is the free-fermion energy,
and $a_{s}$ is the $s$-wave scattering length coming from the renormalization
of the coupling constant~\cite{deMelo1993}:%
\begin{equation}
\frac{1}{g}=\frac{1}{8\pi a_{s}}-\int_{k<K_{0}}\frac{d\mathbf{k}}{\left(
2\pi\right)  ^{3}}\frac{1}{2k^{2}}, \label{g}%
\end{equation}
with the ultraviolet cutoff $K_{0}$. In the case of the model contact
interaction, $g\rightarrow0$ and, correspondingly, the cutoff is set to
$K_{0}\rightarrow\infty$, keeping the integral in equation~(\ref{gap2}) convergent.

The resulting GPDF action is a quadratic form of the modulus $\left(
\lambda_{\mathbf{q},m}\right)  $ and phase $\left(  \theta_{\mathbf{q}%
,m}\right)  $ fluctuations of the pair field and the density fluctuations
$\Phi_{\mathbf{q},m}$:%
\begin{align}
S_{GPDF}  &  =\frac{1}{2}\sum_{\mathbf{q},m}\left(
\begin{array}
[c]{ccc}%
\lambda_{-\mathbf{q},-m} & \theta_{-\mathbf{q},-m} & \Phi_{-\mathbf{q},-m}%
\end{array}
\right) \nonumber\\
&  \times\left(
\begin{array}
[c]{ccc}%
K_{1,1} & K_{1,2} & K_{1,3}\\
-K_{1,2} & K_{2,2} & K_{2,3}\\
K_{1,3} & -K_{2,3} & K_{3,3}%
\end{array}
\right)  \left(
\begin{array}
[c]{c}%
\lambda_{\mathbf{q},m}\\
\theta_{\mathbf{q},m}\\
\Phi_{\mathbf{q},m}%
\end{array}
\right)  . \label{SGPDF4}%
\end{align}
{Here,} $\mathbf{q}$ is the momentum, and $m$ is the Matsubara index of the
Fourier representation of the bosonic fields. The coefficients $K_{i,j}$
constitute the matrix of the inverse GPDF propagator. They are explicitly
given by (for details, see~\cite{Plasmons,Plasmons2,Castin}):{\small
\begin{align}
K_{1,1}\left(  \mathbf{q},i\Omega_{m}\right)   &  =-\frac{1}{8\pi a_{s}}%
+\int\frac{d\mathbf{k}}{\left(  2\pi\right)  ^{3}}\left\{  \frac{1}{2k^{2}%
}+\frac{\tanh\left(  \beta E_{\mathbf{k}}/2\right)  }{4E_{\mathbf{k}%
}E_{\mathbf{k}+\mathbf{q}}}\right. \nonumber\\
&  \times\left[  \left(  \xi_{\mathbf{k}}\xi_{\mathbf{k}+\mathbf{q}%
}+E_{\mathbf{k}}E_{\mathbf{k}+\mathbf{q}}-\Delta^{2}\right)  \left(  \frac
{1}{i\Omega_{m}-E_{\mathbf{k}}-E_{\mathbf{k}+\mathbf{q}}}-\frac{1}{i\Omega
_{m}+E_{\mathbf{k}}+E_{\mathbf{k}+\mathbf{q}}}\right)  \right. \nonumber\\
&  \left.  \left.  +\left(  \xi_{\mathbf{k}}\xi_{\mathbf{k}+\mathbf{q}%
}-E_{\mathbf{k}}E_{\mathbf{k}+\mathbf{q}}-\Delta^{2}\right)  \left(  \frac
{1}{i\Omega_{m}-E_{\mathbf{k}+\mathbf{q}}+E_{\mathbf{k}}}-\frac{1}{i\Omega
_{m}-E_{\mathbf{k}}+E_{\mathbf{k}+\mathbf{q}}}\right)  \right]  \right\}  ,
\label{K11}%
\end{align}%
\begin{align}
K_{2,2}\left(  \mathbf{q},i\Omega_{m}\right)   &  =-\frac{1}{8\pi a_{s}}%
+\int\frac{d\mathbf{k}}{\left(  2\pi\right)  ^{3}}\left\{  \frac{1}{2k^{2}%
}+\frac{\tanh\left(  \beta E_{\mathbf{k}}/2\right)  }{4E_{\mathbf{k}%
}E_{\mathbf{k}+\mathbf{q}}}\right. \nonumber\\
&  \times\left[  \left(  \xi_{\mathbf{k}}\xi_{\mathbf{k}+\mathbf{q}%
}+E_{\mathbf{k}}E_{\mathbf{k}+\mathbf{q}}+\Delta^{2}\right)  \left(  \frac
{1}{i\Omega_{m}-E_{\mathbf{k}}-E_{\mathbf{k}+\mathbf{q}}}-\frac{1}{i\Omega
_{m}+E_{\mathbf{k}}+E_{\mathbf{k}+\mathbf{q}}}\right)  \right. \nonumber\\
&  \left.  \left.  +\left(  \xi_{\mathbf{k}}\xi_{\mathbf{k}+\mathbf{q}%
}-E_{\mathbf{k}}E_{\mathbf{k}+\mathbf{q}}+\Delta^{2}\right)  \left(  \frac
{1}{i\Omega_{m}-E_{\mathbf{k}+\mathbf{q}}+E_{\mathbf{k}}}-\frac{1}{i\Omega
_{m}-E_{\mathbf{k}}+E_{\mathbf{k}+\mathbf{q}}}\right)  \right]  \right\}  ,
\label{K22}%
\end{align}
}
\begin{align}
K_{1,2}\left(  \mathbf{q},i\Omega_{m}\right)   &  =i\int\frac{d\mathbf{k}%
}{\left(  2\pi\right)  ^{3}}\frac{\tanh\left(  \beta E_{\mathbf{k}}/2\right)
}{4E_{\mathbf{k}}E_{\mathbf{k}+\mathbf{q}}}\nonumber\\
&  \times\left[  \left(  \xi_{\mathbf{k}}E_{\mathbf{k}+\mathbf{q}%
}+E_{\mathbf{k}}\xi_{\mathbf{k}+\mathbf{q}}\right)  \left(  \frac{1}%
{i\Omega_{m}-E_{\mathbf{k}}-E_{\mathbf{k}+\mathbf{q}}}+\frac{1}{i\Omega
_{m}+E_{\mathbf{k}}+E_{\mathbf{k}+\mathbf{q}}}\right)  \right. \nonumber\\
&  \left.  +\left(  \xi_{\mathbf{k}}E_{\mathbf{k}+\mathbf{q}}-E_{\mathbf{k}%
}\xi_{\mathbf{k}+\mathbf{q}}\right)  \left(  \frac{1}{i\Omega_{m}%
-E_{\mathbf{k}+\mathbf{q}}+E_{\mathbf{k}}}+\frac{1}{i\Omega_{m}-E_{\mathbf{k}%
}+E_{\mathbf{k}+\mathbf{q}}}\right)  \right]  , \label{K12}%
\end{align}%
\begin{align}
K_{1,3}\left(  \mathbf{q},i\Omega_{m}\right)   &  =-i\sqrt{2\alpha_{0}}%
\Delta\int\frac{d\mathbf{k}}{\left(  2\pi\right)  ^{3}}\frac{\tanh\left(
\beta E_{\mathbf{k}}/2\right)  }{4E_{\mathbf{k}}E_{\mathbf{k}+\mathbf{q}}%
}\left(  \xi_{\mathbf{k}}+\xi_{\mathbf{k}+\mathbf{q}}\right) \nonumber\\
&  \times\left(  \frac{1}{i\Omega_{m}-E_{\mathbf{k}}-E_{\mathbf{k}+\mathbf{q}%
}}-\frac{1}{i\Omega_{m}+E_{\mathbf{k}}+E_{\mathbf{k}+\mathbf{q}}}\right)
\nonumber\\
&  \left.  +\frac{1}{i\Omega_{m}+E_{\mathbf{k}}-E_{\mathbf{k}+\mathbf{q}}%
}-\frac{1}{i\Omega_{m}-E_{\mathbf{k}}+E_{\mathbf{k}+\mathbf{q}}}\right)  ,
\label{K13}%
\end{align}%
\begin{align}
K_{2,3}\left(  \mathbf{q},i\Omega_{m}\right)   &  =-\sqrt{2\alpha_{0}}%
\Delta\int\frac{d\mathbf{k}}{\left(  2\pi\right)  ^{3}}\frac{\tanh\left(
\beta E_{\mathbf{k}}/2\right)  }{4E_{\mathbf{k}}E_{\mathbf{k}+\mathbf{q}}%
}\nonumber\\
&  \times\left[  \left(  E_{\mathbf{k}+\mathbf{q}}+E_{\mathbf{k}}\right)
\left(  \frac{1}{i\Omega_{m}-E_{\mathbf{k}}-E_{\mathbf{k}+\mathbf{q}}}%
+\frac{1}{i\Omega_{m}+E_{\mathbf{k}}+E_{\mathbf{k}+\mathbf{q}}}\right)
\right. \nonumber\\
&  \left.  +\left(  E_{\mathbf{k}+\mathbf{q}}-E_{\mathbf{k}}\right)  \left(
\frac{1}{i\Omega_{m}+E_{\mathbf{k}}-E_{\mathbf{k}+\mathbf{q}}}+\frac
{1}{i\Omega_{m}-E_{\mathbf{k}}+E_{\mathbf{k}+\mathbf{q}}}\right)  \right]  ,
\label{K23}%
\end{align}%
\begin{align}
K_{3,3}\left(  \mathbf{q},i\Omega_{m}\right)   &  =\frac{q^{2}}{4\pi}%
-\alpha_{0}\int\frac{d\mathbf{k}}{\left(  2\pi\right)  ^{3}}\frac{\tanh\left(
\beta E_{\mathbf{k}}/2\right)  }{2E_{\mathbf{k}}E_{\mathbf{k}+\mathbf{q}}%
}\nonumber\\
&  \times\left(  \frac{E_{\mathbf{k}}E_{\mathbf{k}+\mathbf{q}}-\xi
_{\mathbf{k}}\xi_{\mathbf{k}+\mathbf{q}}+\Delta^{2}}{i\Omega_{m}%
-E_{\mathbf{k}}-E_{\mathbf{k}+\mathbf{q}}}+\frac{E_{\mathbf{k}}E_{\mathbf{k}%
+\mathbf{q}}+\xi_{\mathbf{k}}\xi_{\mathbf{k}+\mathbf{q}}-\Delta^{2}}%
{i\Omega_{m}-E_{\mathbf{k}}+E_{\mathbf{k}+\mathbf{q}}}\right. \nonumber\\
&  \left.  -\frac{E_{\mathbf{k}}E_{\mathbf{k}+\mathbf{q}}+\xi_{\mathbf{k}}%
\xi_{\mathbf{k}+\mathbf{q}}-\Delta^{2}}{i\Omega_{m}+E_{\mathbf{k}%
}-E_{\mathbf{k}+\mathbf{q}}}-\frac{E_{\mathbf{k}}E_{\mathbf{k}+\mathbf{q}}%
-\xi_{\mathbf{k}}\xi_{\mathbf{k}+\mathbf{q}}+\Delta^{2}}{i\Omega
_{m}+E_{\mathbf{k}}+E_{\mathbf{k}+\mathbf{q}}}\right)  . \label{K33}%
\end{align}
{The} other matrix elements can be written down using the symmetry relations,%
\begin{align}
K_{2,1}\left(  \mathbf{q},i\Omega_{m}\right)   &  =-K_{1,2}\left(
\mathbf{q},i\Omega_{m}\right)  ,\label{K21}\\
K_{3,1}\left(  \mathbf{q},i\Omega_{m}\right)   &  =K_{1,3}\left(
\mathbf{q},i\Omega_{m}\right)  ,\\
K_{3,2}\left(  \mathbf{q},i\Omega_{m}\right)   &  =-K_{2,3}\left(
\mathbf{q},i\Omega_{m}\right)
\end{align}

As the matrix elements of the inverse GPDF propagator $K_{1,3}$ and $K_{2,3}$
are proportional to the gap $\Delta$, the pair and density fields become
decoupled at $T=T_{c}$. For lower temperatures, however, this coupling can
influence the spectra of collective modes. Moreover, as shown below, even in a
relatively close vicinity of the transition temperature, at $T=0.99T_{c}$, the
coupling of the pair and density fields leads to clearly observable
consequences, such as an appearance of new collective modes.

Frequencies and damping factors of collective excitations are determined in a
mutually consistent way through roots of the determinant of the inverse GPDF
propagator for the complex frequency argument $z=\omega+i\gamma$,%
\begin{equation}
\det\mathbb{K}\left(  z\right)  =0, \label{DispEq}%
\end{equation}
or, in other words, as complex poles of the GPDF propagator. In order to get
these poles, the propagator is analytically continued through the branch cut
at the real axis (for more details, see Ref.~\cite{Plasmons2}). This method is
well established and widely used, particularly for Green's functions
\cite{Nozieres}, for the dielectric function~\cite{Katsnelson}, for collective
excitations of ultracold Fermi gases
\cite{PB-PRL,AllModes-PRA}.

The matrix elements $K_{j,k}\left(  \mathbf{q},z\right)  $ contain several
angular points at the branch cut. The analytic continuation is therefore
possible through several intervals between these points, as analyzed in detail
in Ref.~\cite{AllModes-PRA}. Different intervals can reveal different branches
of collective excitations.

\begin{figure}[h]
\includegraphics[
height=2.5002in,
width=2.8954in
]{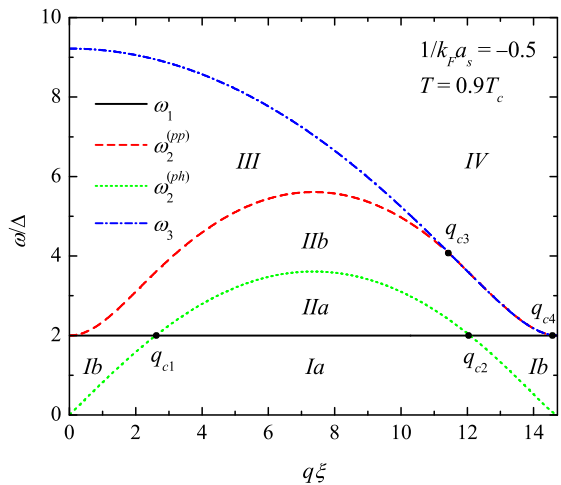}\caption{(Adapted from Ref. \cite{AllModes-PRA}.) Angular-point
frequencies of the GPDF matrix elements for $1/k_{F}a_{s}=0.5$ and
$T=0.9T_{c}$. The areas between curves determine intervals for the analytic
continuation. The momentum is multiplied by the coherence length $\xi\equiv
v_{F}/T_{c}$, where $v_{F}$ is the Fermi velocity ($v_{F}=2$ in the
present~units).}%
\label{fig:Bounds}%
\end{figure}

Bounds of intervals for the analytic continuation shown in
Figure~\ref{fig:Bounds} (and considered in detail in Ref.~\cite{AllModes-PRA})
influence both the analytic solutions of the dispersion equation and the
spectral weight functions, where they are explicitly manifested as shown
below. In the figure, the frequency $\omega_{1}$ is the pair-breaking
continuum edge. The frequencies $\omega_{2}^{\left(  pp\right)  }$ and
$\omega_{2}^{\left(  ph\right)  }$ correspond to bounds of channels for the
particle--particle and particle--hole scattering processes, respectively, and
$\omega_{3}=2\sqrt{\left(  \mu-q^{2}/4\right)  ^{2}+\Delta^{2}}$ is the energy
of the BCS pair $E_{\mathbf{k}-\mathbf{q}/2}+E_{\mathbf{k}+\mathbf{q}/2}$ at
zero fermion momentum $\mathbf{k}$. The set of these bounds is determined by a
change in the configuration of the resonant wave vectors for any of the two
resonance conditions, $\omega=E_{\mathbf{k}-\mathbf{q}/2}+E_{\mathbf{k}%
+\mathbf{q}/2}$ for the particle--particle scattering channel and
$\omega=\left\vert E_{\mathbf{k}-\mathbf{q}/2}-E_{\mathbf{k}+\mathbf{q}%
/2}\right\vert $ for the particle--hole channel. The values $q_{cj}$ indicate
bounds of intervals of momentum where different branches of collective
excitations can exist. For completeness, we can also compute spectral weight functions
for the phase--phase and modulus--modulus correlation functions of the pair
field and for the density--density correlation function. In the general form,
they are reported in Ref.~\cite{Plasmons2}. Here, we approximate them assuming
the limit $\omega_{p}\rightarrow\infty$ for the plasma frequency. This
approximation is relevant for low-lying collective excitations, and reads%
\begin{align}
\lim_{\alpha_{0}\rightarrow\infty}\chi_{\lambda\lambda}\left(  \mathbf{q}%
,\omega\right)   &  =\frac{1}{\pi}\operatorname{Im}\frac{K_{2,2}\left(
\mathbf{q},\omega+i0^{+}\right)  \tilde{K}_{3,3}\left(  \mathbf{q}%
,\omega+i0^{+}\right)  +\tilde{K}_{2,3}^{2}\left(  \mathbf{q},\omega
+i0^{+}\right)  }{\det\mathbb{\tilde{K}}\left(  \mathbf{q},\omega
+i0^{+}\right)  },\label{XLLb}\\
\lim_{\alpha_{0}\rightarrow\infty}\chi_{\theta\theta}\left(  \mathbf{q}%
,\omega\right)   &  =\frac{1}{\pi}\operatorname{Im}\frac{K_{1,1}\left(
\mathbf{q},\omega+i0^{+}\right)  \tilde{K}_{3,3}\left(  \mathbf{q}%
,\omega+i0^{+}\right)  -\tilde{K}_{1,3}^{2}\left(  \mathbf{q},\omega
+i0^{+}\right)  }{\det\mathbb{\tilde{K}}\left(  \mathbf{q},\omega
+i0^{+}\right)  },\label{XYYb}\\
\lim_{\alpha_{0}\rightarrow\infty}\chi_{\rho\rho}\left(  \mathbf{q}%
,\omega\right)   &  =-\frac{q^{4}}{16\pi^{3}}\operatorname{Im}\left(
\frac{\det\mathbb{K}_{GPF}\left(  \mathbf{q},\omega+i0^{+}\right)  }%
{\det\mathbb{\tilde{K}}\left(  \mathbf{q},\omega+i0^{+}\right)  }\right)  ,
\label{RF1b}%
\end{align}
with the rescaled matrix elements and the determinant
\begin{equation}
\tilde{K}_{1,3}=\frac{1}{\sqrt{\alpha_{0}}}K_{1,3},\quad\tilde{K}_{2,3}%
=\frac{1}{\sqrt{\alpha_{0}}}K_{2,3},\quad\tilde{K}_{3,3}=\frac{1}{\alpha_{0}%
}\left(  K_{3,3}-\frac{q^{2}}{4\pi}\right)  , \label{mels}%
\end{equation}%
\begin{equation}
\det\mathbb{\tilde{K}}=\lim_{\alpha_{0}\rightarrow\infty}\left(  \frac
{1}{\alpha_{0}}\det\mathbb{K}\right)  , \label{limK}%
\end{equation}
which tend to finite and nonzero limiting values when $\alpha_{0}%
\rightarrow+\infty$.

In the limit of large $\omega_{p}$, the contribution of the low-lying
collective excitations in the density--density spectral weight is extremely
small. Moreover, in this limit, the contribution of the density oscillations
to the low-lying excitations must be relatively small. Therefore, we use the
pair--field spectral weight functions in the subsequent numeric results. The
phase--phase and modulus--modulus spectral weight functions are useful to
clarify the physical sense of different modes and to see whether they are
provided by oscillations of the modulus or the phase of the pair field. Thus,
the spectrum of collective excitations is determined below in two
complementary ways, (1) as peaks of spectral weight functions, and (2) finding
a solution for complex poles of the GPDF propagator analytically continued to
the lower half-plane of the complex frequency.

The random-phase and Gaussian fluctuation approximations are well
substantiated and widely used in the whole BCS--BEC crossover, as long as we
stay under the assumption that collective excitations are \emph{{small}}
~(harmonic) oscillations about the uniform background. The special case when
higher-order terms in the power of fluctuations are important is considered in
Ref.~\cite{Nonquad}. The other reason for a quantitative inaccuracy of the
present approach can follow from the mean-field results for the uniform gap
and the chemical potential. This does not influence the qualitative behavior
of collective excitations as discussed in Refs.~\cite{PB-PRL,AllModes-PRA}.

\section{Spectra of Collective Excitations}

There is a significant difference between the behavior of collective
excitations in neutral and charged fermionic superfluids in the frequency
range of the order of the pair-breaking threshold. In a neutral superfluid
Fermi gas, there exists the gapless (Anderson--Bogoliubov, Goldstone)
collective excitation branch provided by the phase response of the pair field,
and the gapped (pair-breaking, Higgs) branch constituted by the modulus
response. The Anderson--Bogoliubov mode turns to the gapped plasma mode due to
the Coulomb interaction, and the Goldstone phase mode is thus suppressed
\cite{Anderson1958}. When the temperature rises, approaching the vicinity of
the transition temperature $T_{c}$, a gapless mode appears again due to the
presence of the normal fraction, as described by Carlson and Goldman
\cite{Carlson}.

\begin{figure}[h]
\includegraphics[
height=3.96781in,
width=4.83824in
]{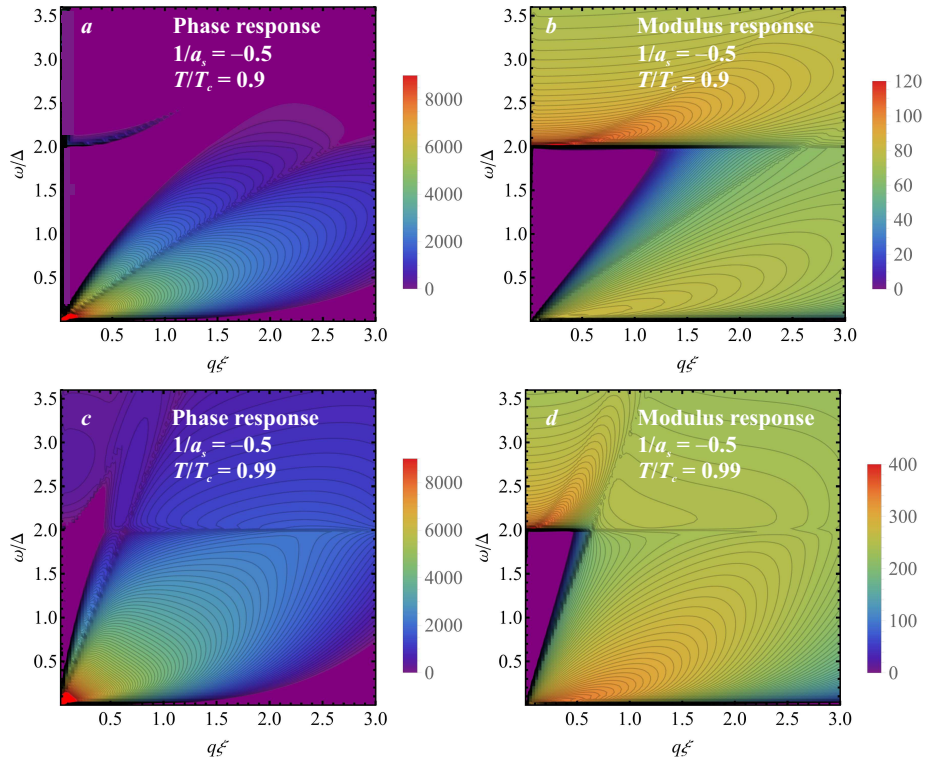}\caption{Contour plots of the spectral weight functions at
$T=0.9T_{c}$ (\textbf{a},\textbf{b}) and at $T=0.99T_{c}$ (\textbf{c}%
,\textbf{d}) for the phase (\textbf{a},\textbf{c}) and modulus (\textbf{b}%
,\textbf{d}) response of a charged Fermi superfluid with the inverse $s$-wave
scattering length $1/k_{F}a_{s}=-0.5$. The momentum is measured in units of
$1/\xi$, where $\xi\equiv v_{F}/T_{c}$ with the Fermi velocity $v_{F}=\hbar
k_{F}/m$. The clipping area above the plot range for the spectral weights is
shown by red color.}%
\label{fig:Contours1}%
\end{figure}

In the BCS--BEC crossover, the strong coupling is favorable for the
manifestation of the gapless collective excitation branch with respect to that
in the far BCS regime. This extended range of survival for gapless excitations
is revealed in the contour plots of the spectral weight functions for the
phase and modulus response of a charged superfluid in
Figure~\ref{fig:Contours1}.~In the far BCS limit~\cite{Takada1997}, the
gapless mode disappears below $T/T_{c}\approx0.9$, while it is rather strongly
expressed at that temperature even in the moderate BCS case with $1/k_{F}%
a_{s}=-0.5$, and moreover at higher coupling strengths. The logarithmic scale
is used for the better visibility of the whole spectrum. Additionally, we
partly clipped the plot range for the spectral weights of the phase response
in Figure~\ref{fig:Contours1}{a},{c}. The clipping areas are shown in red.

Comparing the phase and modulus spectral weight functions to each other, it
can be seen that gapless collective excitations in the BCS--BEC crossover
appear in both phase and modulus spectral weight functions such that they are~
provided by both the modulus and phase response as distinct from the far BCS
limit $1/a_{s}\rightarrow-\infty$~\cite{Takada1997}, where they are
constituted by phase excitations only. The spectral weight functions show the
existence of more than one gapless branch. One of them contains a significant
contribution of amplitude excitations, with an increasing relative weight when
$T$ approaches $T_{c}$, as can be seen from Figure~\ref{fig:Contours1}{b},{d}.
The other gapless branches are not visible in the modulus spectral weight
functions. Consequently, they can be attributed to the phase response. We can
see a splitting of these collective excitations to the branches above and
below the particle--hole angular point frequency $\omega_{2}^{\left(
ph\right)  }$, which is revealed as a kink in the contour plots.

The pair-breaking collective excitation branch is less affected by the Coulomb
interaction and exists both in superconductors and in neutral Fermi
superfluids. The pair-breaking Higgs modes do not contain a sufficiently
resolvable phase component and, therefore, are not visible in the spectral
weight functions for the phase response (Figure~\ref{fig:Contours1}{a},{c}).
They are provided by amplitude oscillations in both neutral and charged superfluids.

\begin{figure}[h]
\includegraphics[
height=3.96781in,
width=4.83824in
]{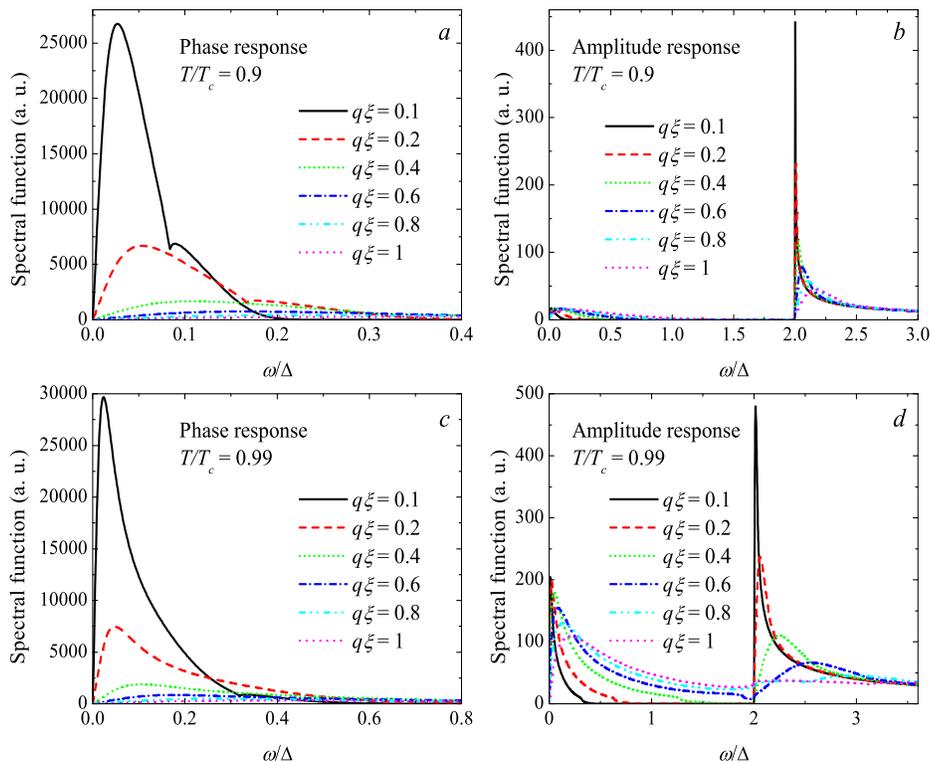}\caption{Spectral weight functions of the phase
(\textbf{a},\textbf{c}) and amplitude (\textbf{b},\textbf{d}) pair field
response at $T/T_{c}=0.9$ (\textbf{a},\textbf{b}) and $T/T_{c}=0.99$
(\textbf{c},\textbf{d}) for several values of the field momentum $q$.}%
\label{fig:SFunctions1D}%
\end{figure}

The one-dimensional slices of the spectral weight functions at selected values
of the pair field momentum $q$ shown in Figure~\ref{fig:SFunctions1D} give us
even more transparent picture of the behavior of different collective
excitation branches. The kink in the spectral weight functions of the phase
response manifested in Figure~\ref{fig:SFunctions1D}a,c at $\omega=\omega
_{2}^{\left(  ph\right)  }$ apparently indicates the existence of two gapless
modes related to phase excitations. They are not well resolved from each other
because in the temperature range where gapless modes exist they exhibit
significant damping. Consequently, the two peaks in
Figure~\ref{fig:SFunctions1D}a,c substantially overlap.

According to the spectral weight functions in comparison with the results of
Ref. \cite{AllModes-PRA}, the gapless mode splits to more branches in charged
superfluids/superconductors than in neutral superfluids. It is not {a priori}
clear whether or not this splitting is accompanied by an appearance of more
than one complex poles in the analytic solution. Below, we show that the
splitting is indeed consistent with the behavior of the poles of the propagator.

The spectral weight functions for the amplitude response in Figure
\ref{fig:SFunctions1D}b,d show peaks attributed to the gapped Higgs modes and
gapless modes. The Higgs mode frequencies are pinned to the pair-breaking
continuum edge and tend to $2\Delta$ at the small momentum. Their behavior in
charged superfluids is only slightly influenced by plasma oscillations at
$\omega_{p}\gg\Delta$, contrary to the case of small plasma frequencies
considered in Ref.~\cite{Plasmons2}, where Higgs and plasma modes can be in resonance.

Besides the Higgs modes, the amplitude response exhibits the lowest-energy
gapless excitation branch, which shows a distinctive dissipative behavior. Its
response magnitude increases when moving close toward $T_{c}$.

\begin{figure}[h]
\includegraphics[
height=3.3347in,
width=5.0021in
]{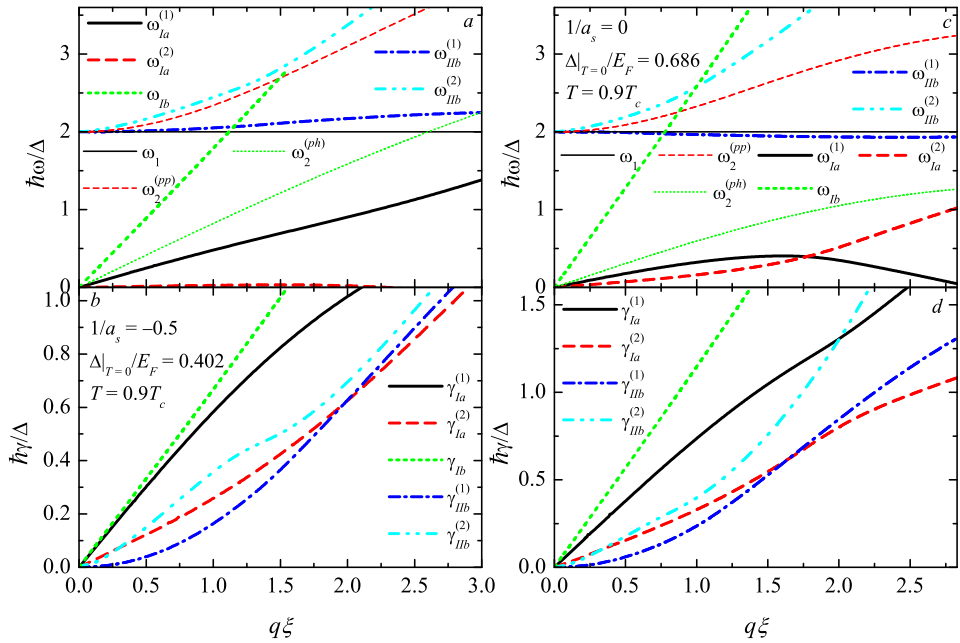}\caption{Heavy solid, dashed and dotted curves: eigenfrequency
(\textbf{a},\textbf{c}) and damping factor (\textbf{b},\textbf{d}) of
low-lying collective excitations for the inverse scattering length
$1/k_{F}a_{s}=-0.5$ (\textbf{a},\textbf{b}) and $1/k_{F}a_{s}=0$
(\textbf{c},\textbf{d}) at the relative temperature $T/T_{c}=0.9$. Thin solid,
dashed and dotted curves show angular-point frequencies, which indicate bounds
of energy intervals for different scattering processes.}%
\label{fig:DispT09}%
\end{figure}

Next, we analyze complex eigenfrequencies given by roots of the dispersion
equation~(\ref{DispEq}). The solution based on the analytic continuation of
the GPDF propagator exhibits more complex poles in the BCS--BEC crossover with
respect to the far BCS limit, in accordance with the spectral weight functions
shown in Figure~\ref{fig:Contours1}. Only a part of them are really physically
significant, namely those that have relatively small damping rates, because in
general, complex poles of Green's functions can be convincingly attributed to
elementary excitations only at sufficiently small damping~\cite{AGDBook}.
Figure \ref{fig:DispT09} shows the momentum dependence of eigenfrequencies and
damping factors in the moderate BCS regime for $T=0.9T_{c}$ and for two values
of the inverse scattering length, $1/k_{F}a_{s}=-0.5$, which corresponds to
Figure~\ref{fig:Contours1}, and the unitarity regime with $1/k_{F}a_{s}=0$.

Even at small damping, the mutually consistent calculation of eigenfrequencies
and damping factors is more rigorous than the perturbation approach, where the
eigenfrequency is determined for the zero damping, and the damping rate is
then calculated within the lowest-order perturbation theory. The perturbation
approach can give solutions such as an upper mode at $\omega\approx2\Delta$
for phase excitations, which may be unphysical, as mentioned in Ref.
\cite{Takada1997}.

We show in Figure~\ref{fig:DispT09} the most representative (with
non-negligible spectral weights) branches of gapless collective excitations.
These roots of the dispersion equation (\ref{DispEq}) are generated by the
analytic continuation of the GPDF propagator through the intervals $Ia$ and
$Ib$ from Figure~\ref{fig:Bounds}. In agreement with the spectral weight
functions, we can see three of the most representative complex poles with the
eigenfrequencies $\left(  \omega_{Ia}^{\left(  1\right)  },\omega
_{Ia}^{\left(  2\right)  }\right)  <\omega_{2}^{\left(  ph\right)  }$ and
$\omega_{Ib}>\omega_{2}^{\left(  ph\right)  }$. The latter pole does not exist
in the solution for neutral superfluids~\cite{AllModes-PRA}. Therefore, the
appearance of this gapless mode is provided by the interaction of the phase
and plasma collective excitations. These two upper-frequency gapless modes
have the sound-like (linear) dispersion at the small momentum. This looks like
the splitting of the phononic-like Carlson0-Goldman branch to two
phononic-like collective excitation modes. Figure~\ref{fig:DispT09}a
demonstrates an example of the upper-frequency gapless mode pole
$z_{Ia}^{\left(  2\right)  }$ existing in the restricted area of momentum
$q<q_{1c}$, as follows from the scheme of intervals for the analytic
continuation (Figure~\ref{fig:Bounds}).

The lower-frequency gapless branch with the eigenfrequency $\omega
_{Ia}^{\left(  2\right)  }$ is an analog of the second gapless mode in neutral
Fermi superfluids~\cite{PB-PRA}. It exists at $T$ sufficiently close to the
transition temperature and disappears at some temperature below $T_{c}$, which
depends on the inverse scattering length, and decreases when rising the
coupling strength. As can be seen from Figure~\ref{fig:Bounds}, this second
mode is close to disappearance at $T=0.9T_{c}$ and $1/a_{s}=-0.5$, but it is
non-vanishing at $1/a_{s}=0$ with the same relative temperature. Moreover, at
unitarity, we can see the crossing of the solutions $\omega_{Ia}^{\left(
1\right)  }$ and $\omega_{Ia}^{\left(  2\right)  }$ at $q\xi\approx1.6$. Thus,
at a larger momentum, $\omega_{Ia}^{\left(  2\right)  }$ becomes the upper
gapless mode of the two. This crossing of frequencies is accompanied by the
avoided crossing of the damping factors so that complex roots are not crossed.
This behavior of complex eigenfrequencies is qualitatively the same as that
for neutral Fermi superfluids, discussed in Refs.~\cite{AllModes-PRA,PB-PRA}.

The pair-breaking collective excitation branch in the BCS--BEC crossover and
at $T\neq0$ splits to two branches $\omega_{IIb}^{\left(  1\right)  }$ and
$\omega_{IIb}^{\left(  2\right)  }$, similarly to the pair-breaking mode in a
neutral fermionic superfluid~\cite{AllModes-PRA}. The first solution
$\omega_{IIb}^{\left(  1\right)  }$ dominates at a small momentum with respect
to $\omega_{IIb}^{\left(  2\right)  }$ due to a smaller damping. It exhibits a
negative dispersion approaching a minimum at some finite $q$. This behavior
has an analog with the negative dispersion of the plasma mode~\cite{Plasmons}
found when the plasma frequency is low and comparable with the gap.

\begin{figure}[h]
\includegraphics[
height=3.3347in,
width=4.99in
]{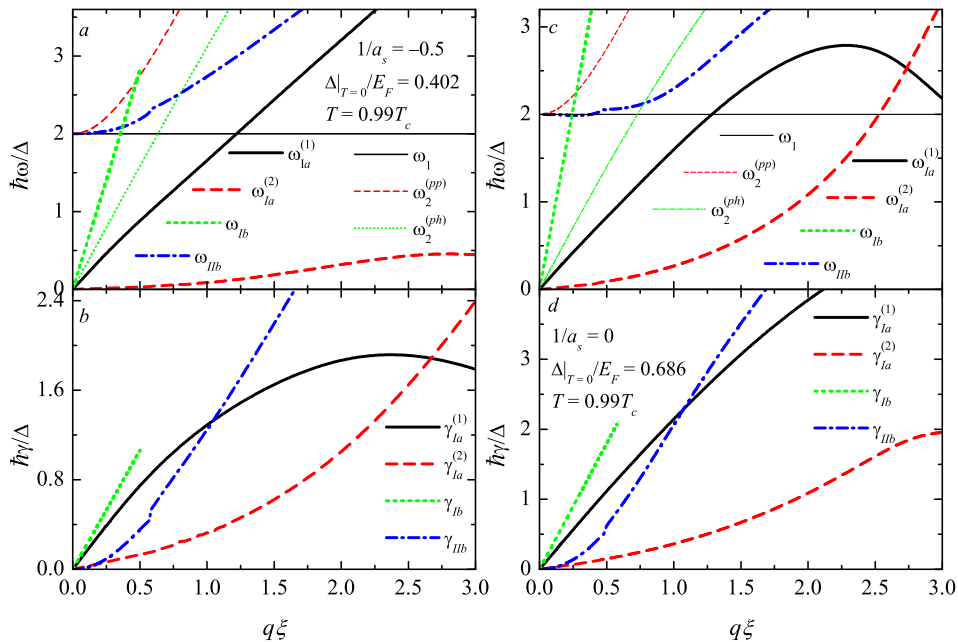}\caption{Eigenfrequency (\textbf{a},\textbf{c}) and damping factor
(\textbf{b},\textbf{d}) of low-lying collective excitations for the inverse
scattering length $1/k_{F}a_{s}=-0.5$ (\textbf{a},\textbf{b}) and
$1/k_{F}a_{s}=0$ (\textbf{c},\textbf{d}) at the relative temperature
$T/T_{c}=0.99$. The notations are the same as in Figure~\ref{fig:DispT09}.}%
\label{fig:DispT99}%
\end{figure}

The gapless mode frequencies $\omega_{Ia}^{\left(  1\right)  }$ and
$\omega_{Ib}$ expand to the continuum when $T$ is close to $T_{c}$, as shown
in Figure~\ref{fig:DispT99}, similarly to the phononic-like mode in a neutral
superfluid~\cite{AllModes-PRA}. This mode frequency demonstrates an avoided
crossing with the frequency of the pair-breaking mode frequency $\omega_{IIb}$
above the threshold, accompanied by the crossing of their damping factors. The
other gapless mode $z_{Ib}$ exhibits crossing with the Higgs mode for
frequencies but no crossing for damping factors. The two gapless branches
$\omega_{Ia}^{\left(  1\right)  }$ and $\omega_{Ia}^{\left(  2\right)  }$
manifest the crossing in the BCS regime at $1/k_{F}a_{s}=-0.5$, which is
changed to the avoided crossing at a strong coupling with $1/k_{F}a_{s}=0$.
Correspondingly, the damping factors $\gamma_{Ia}^{\left(  1\right)  }$ and
$\gamma_{Ia}^{\left(  2\right)  }$ are crossed in the BCS regime and exhibit
an avoided crossing at unitarity. The aforesaid behavior of complex poles is
qualitatively similar to that found for collective excitations in a neutral
superfluid, where they behave as particles interacting with each other via a
repulsive potential.

Comparing the solutions of the dispersion equation (\ref{DispEq})
with the peaks of the spectral weight functions in Figure~\ref{fig:Contours1},
it can be seen that the gapless modes with the frequencies $\omega
_{Ia}^{\left(  1\right)  }$ and $\omega_{Ib}$ are related to the phase
response because there are no visible fingerprints of these modes in the
modulus spectral weight functions. The other gapless mode $\omega
_{Ia}^{\left(  2\right)  }$ is distinctively manifested in the spectral weight
functions of the modulus response, and consequently contains a significant
part of the amplitude oscillations. When the temperature shifts toward $T_{c}%
$, the relative weight of this (lowest energy at small $q$) gapless mode in
the modulus response increases with respect to the relative weight of the
Higgs mode, while the relative weights of the gapless modes in the phase
response vary only slightly.

The dispersion of the gapless mode $\omega_{Ia}^{\left(  2\right)  }$ is
linear at a sufficiently small momentum. When the momentum increases further,
we can observe a range of $q$, where $\omega_{Ia}^{\left(  2\right)  }\left(
q\right)  $ is a convex function. When $T$ approaches a vicinity of $T_{c}$,
the interval of linearity gradually shrinks so that the gapless branch
$\omega_{Ia}^{\left(  2\right)  }$ apparently tends to the mode with a
quadratic dispersion, similarly to the analogous mode in a neutral superfluid
\cite{deMelo1993,AllModes-PRA}.

In general, the real part of a complex pole is interpreted as the
eigenfrequency of the collective mode, and the imaginary part determines the
damping factor, which is the inverse lifetime of the mode. This physical
picture is adequate at relatively small damping factors, when complex poles
are well-determined quasiparticle excitations. For stronger damping, this
understanding gradually becomes rather conventional. Nevertheless, {we can
observe this about collective excitations}
~even when damping is not small, just describing the obtained complex poles as
they appear.

The broadening of collective excitations hardly can be subdivided into the
quantum and thermal parts precisely. Within the GPDF/RPA methods, they come
together in the matrix elements. The broadening of collective modes at $T=0$
is of course only quantum, and it is nonzero only for the nonzero momentum. At
zero temperature, the gapless modes do not survive in charged superfluids.
Therefore, the purely quantum broadening of Carlson--Goldman excitations
hardly can be observed. As we can see from the obtained results in the
figures, the broadening of all modes increases when rising the temperature.

\section{Conclusions}

We considered collective excitations in superconductors and charged Fermi
superfluids in the case when the plasma frequency substantially exceeds the
gap and pair-breaking threshold, focusing on the low-lying collective modes
with energies of the order of the superconducting/superfluid gap. Within the
present study, we find several low-lying collective modes existing in a
charged superfluid/superconductor. The pair-breaking Higgs branch of
collective excitations survives both in neutral and charged superfluid
systems. It is relatively slightly influenced by the Coulomb interaction,
except for the case of resonance of the plasma excitation branch with the
pair-breaking threshold considered in Refs.~\cite{Plasmons,Plasmons2}. In the
BCS--BEC crossover, we found more than one pair-breaking collective excitation~branch.

When the temperature rises toward $T_{c}$, gapless collective excitation
branches can appear in the BCS--BEC crossover. The physical origin of these
modes is the presence of a normal fluid fraction when $T$ is sufficiently
close to $T_{c}$, the same reason as for Carlson--Goldman excitations
\cite{Carlson}. There can appear several gapless modes, distinct from a single
mode in the far BCS limit. At the small momentum, one of the gapless modes
with a linear dispersion can be attributed to the Carlson--Goldman collective
mode as in Ref.~\cite{Takada1997}. The other gapless excitation branch, which
has a lower energy at small momentum, contains a resolvable part of the
modulus response. Despite the amplitude contribution to this mode, it is quite
different from the pair-breaking Higgs modes, which are gapped and pinned to
the pair-breaking threshold. When $T$ approaches $T_{c}$, this mode exhibits a
quadratic dispersion and is analogous to the excitation branch for a neutral
Fermi superfluid described in Ref.~\cite{deMelo1993}. Finally, the other
gapless mode, which is sound-like at a small momentum, appears in the BCS--BEC
crossover for charged superfluids, being new with respect to the previously
considered neutral superfluids.

The spectral weight function for the phase response reveals the gapless modes
but does not contain a resolvable contribution of the Higgs modes. On the
contrary, the modulus response is dominated by the Higgs modes and the second
gapless mode without a visible fingerprint of the Carlson--Goldman branch,
which remains, therefore, a phase mode both in the BCS and crossover regimes.

The evolution of low-lying collective excitation spectra as functions of the
interaction strength and on the temperature can reveal a resonant interaction
of different modes, particularly their avoided crossing. Additionally,
multiple branches of gapless and pair-breaking excitations can be observable.
Because the collective excitations analyzed in the present work are common for
superconductors and charged superfluid systems, they are an example of a
bridge between the physics of superconductors and condensed quantum gases, and
can thus represent a particular interest for both theoretical and experimental investigations.



\end{document}